\documentclass[aps,pre,two column,groupedaddress]{revtex4-2}
\usepackage{amsmath, amsfonts, amssymb}
\usepackage{physics}
\usepackage{graphicx}
\graphicspath{{./images}}
\usepackage{caption, subcaption}

\begin{document}


\title{Collective group drift in a PDE-based opinion dynamics model with biased perception kernels}


\author{Christian Koertje$^{1,}$}
\email[]{ckoertje@umass.edu}
\homepage[]{https://ckoertje.github.io/}
\altaffiliation{Department of Physics, University of Massachusetts Amherst, Amherst, MA 01003, USA}
\author{Hiroki Sayama$^{1,2}$}
\affiliation{$^1$Binghamton Center of Complex Systems (CoCo), Binghamton University, Binghamton, New York 13902-6000, USA;}
\affiliation{$^2$Waseda Innovation Lab, Waseda University, Shinjuku, Tokyo 169-8050, Japan}


\date{\today}

\begin{abstract}
In the age of technology, individuals accelerate their biased gathering of information which in turn leads to a population becoming extreme and more polarized. Here we study a partial differential equation model for opinion dynamics that exhibits collective behavior subject to nonlocal interactions. We developed a new interaction kernel function to represent biased information gathering. Through a linear stability analysis, we show that biased populations can still form opinionated groups. However, a population that is too heavily biased can no longer come to a consensus, that is, the initial homogeneous mixed state becomes stable. Numerical simulations with biased information gathering show the ability for groups to collectively drift towards one end of the opinion space. This means that a small bias in each individual will collectively lead to groups of individuals becoming extreme together. The characteristic time scale for a groups existence is captured from numerical experiments using the temporal correlation function. Supplementing this, we included a measure of how different each population is after regular time intervals using a form of the Manhattan and Euclidean distance metrics. We conclude by exploring how wall boundary conditions induce pattern formation initially on the most extreme sides of the domain. 
\end{abstract}


\maketitle

\section{Introduction}
The study of pattern formation dynamics in nature reveals to us how a collection of interacting constituents come together to form unique spatial structures and maintain complex spatial dynamics. The development of modeling techniques for natural systems have lead to the modeling of systems like social dynamics in which the constituents are complex human beings \cite{castellano2009statistical,Jusup_2022}. In particular, the dynamics of how people come to an agreement or disagreement to form groups is very similar to the processes of particle aggregation seen in nature. An understanding of structural dynamics of our society is beneficial for predicting social trends of large behavioral groups and in engineering solutions to prevent/alleviate formation of groups with extreme beliefs.

Extremism can arise in public opinions from social interactions and from consuming extreme media \cite{hopp2020people}. When possible choices become extreme, deliberation of those opinions leads to increased group polarization \cite{glaeser2009extremism}. In social interactions, individuals copy each other's opinions which leads to a growth of extremism \cite{galam2005heterogeneous}. Uncertainty in deliberation of opinions has been studied to show the impact of which contact with extreme individuals can have on open-minded individuals \cite{deffuant2002can}. The conclusion here is that some populations tend to be drawn to exotic/eccentric opinions. Studies on social media analysis capture how post interactions are more prevalent on those that display eccentric ideology \cite{pandey2023generation}.

In general, the expected behavior of any model of opinion formation is the dynamic separation of a mixed or uniform state into two or more distinct groups \cite{bednar2021polarization}. Many established methods for modeling this take a discrete approach \cite{castellano2009statistical,Jusup_2022}. The Voter Model, for example, is a pairwise interaction model often studied on a social network with discrete opinion values. This is a convenient approach under the assumption that opinions take on a discrete number of possibilities (perhaps a yes-no question). However, when there are exceedingly large possible opinions, or large populations, it makes sense to approach modeling the system with a continuum. 

Some continuous models involve interaction in a spatial domain where interactions are determined only if two individuals are close enough together. Bounded confidence models assume there is no interaction between opinions beyond a certain distance \cite{lorenz2007continuous}. Deffuant et.\ al \cite{deffuant2000mixing} proposed a continuous analog to the Voter Model in which the opinion variable takes on a continuous state, and the interactions depend on the distance between two selected opinions. Another popular bounded confidence model is that of Hegselmann and Krause (HK) \cite{hegselmann2006truth}. The HK model is very similar to the Deffuant model in that interactions are based on a distance threshold. However, instead of pairwise interactions, the HK model deals with an all-to-one interaction. Every opinion inside the range of a selected individual will influence their motion in the opinion space. The goal of the HK model is to represent meetings or rally dynamics in which lots of information is shared before an individual makes a decision. Modifications to the Deffuant and HK models have been made on adaptive networks to capture dynamic social structures \cite{kan2023adaptive,li2023bounded}. Studies of noise in HK dynamics shows how a phase transition occurs depending on the radius of interaction \mbox{\cite{wang2017noisy}}. This is but another examination of when and how groups form in opinionated systems.

Increased access to information through resources like the Internet, web search engines, social media, and now AI chat bots allow people to interact nonlocally by sharing opinions even if the impact is weak \cite{tsvetkova2017understanding}. The Deffuant model can be modified to an all-to-all interaction making it global \cite{ben2003bifurcations}, however the interactions remain binary. Modifications to interaction functions is a method used for studying view points of populations and their reaction to information they gather \mbox{\cite{sabin2020pull}}. At the population distribution level, Sayama \cite{sayama2020enhanced} proposed a partial differential equation (PDE) model based on nonlocal information gathering using an interaction kernel function. Sayama showed that the opinion distance between groups depends on the width of the interaction function. The Sayama model also included diffusion to account for randomness in human behavior. Diffusion is a mechanism that has been studied in the context of human behavior \cite{ben2005opinion}.

Each of the models above qualitatively describe the process of a mixed population separating into distinct groups. The major issues, however, are that they assume symmetric interactions and the resulting groups fall into static equilibrium and remain unchanged forever. This behavior is not representative of real social systems in long temporal scales. For example, political polarization displays particularly interesting behavior \cite{baldassarri2021emergence,levy2019echo,waldrop2021modeling} apart from steady state solutions observed in the models above. We do not observe the growth of extremism or the feature that people and groups change. Bias compromise is a social process not often captured in opinion dynamic models \cite{nguyen2020bias}.

Here we capture collective group motion by introducing the mechanism of bias where a population will interact more strongly to some opinions over others. We use Sayama's PDE-based model focusing on how changes in perception of information impact the dynamics. This is done by introducing a new interaction kernel function subject to a bias parameter that weighs opinions to the right end of the domain more heavily than opinions on the left. For instance, the right end of the space could represent the eccentric opinions and the left end more traditional. The shape of the interaction kernel function goes from odd symmetric to represent a completely unbiased population to even symmetric to represent a population that is heavily biased.

Subject to this new kernel function, we conduct a linear stability analysis to show that the resulting dynamics of a homogeneous perturbed state can lead to stationary pattern formation as well as dynamic traveling waves not previously studied. We also show that a region of stability for the homogeneous state emerges for heavily biased populations. We ran a series of numerical simulations to demonstrate dynamics like collective group drift that also displays many dislocation defects where groups merge. After the patterns formed, we measured the temporal correlation to study how the popularity of opinions may change over time. On average, with some amount of bias, groups shift with different speeds leading to regions in the opinion space becoming less popular. Measurements of the temporal correlation show the time scale at which this can occur. We also studied how the distributions change after regular time intervals using the Manhattan distance (L$^1$-norm) and Euclidean distance (L$^2$-norm) metrics.

Lastly, we explore the effect boundary conditions have on the group formation. The studies above use periodic boundary conditions for ease of simulation, however we are interested in how the structure of the domain may impact the groups that form. We do so by running simulations with Dirichlet boundary conditions. This is to simulate a scenario in which there is an absolute maximum and minimum for how extreme an individual can be.

\section{Methods}
\begin{figure*}[t]
    \centering
    \includegraphics[width=0.99\textwidth]{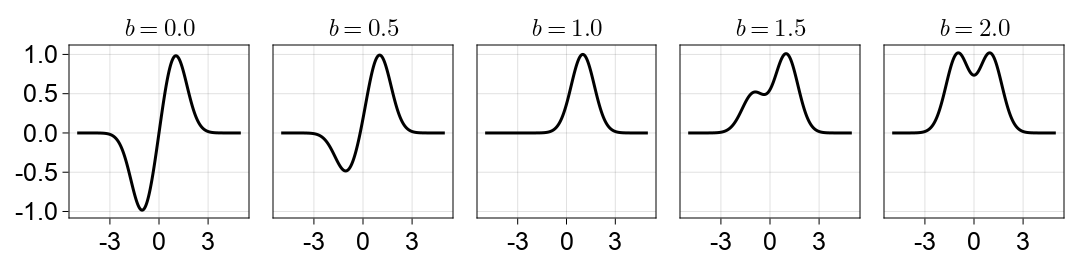}
    \caption{Examples of the functional forms of the perception kernel function Eq.\ \eqref{eq:newKernel} with $\mu = \sigma = 1$. At the extrema, $b=0.0$ the kernel function shape is odd symmetric representing a fully unbiased consumer of information, and $b=2.0$ the population completely repels the left-sided opinion states representing aversion. The horizontal axis can represent relative eccentricity to the point $y=0$. This kernel function is cross-correlated with the opinion distribution to obtain the direction of migration in Eq.\ \eqref{eq:model}.}
    \label{fig:kernels}
\end{figure*}
The model by Sayama \cite{sayama2020enhanced} seeks to account for two key behaviors that govern opinion formation: random choice and an attraction towards popular choices. Modeling this in a continuous spacetime takes the form of a diffusion-migration model
\begin{align}
    \pdv{P}{t} = D_P \nabla^2{P} - c \nabla \cdot \left( P \int_{R} P(x + y, t)g(y) dy \right),
    \label{eq:model}
\end{align}
where $P$ is the population density as a function of opinion space and time, $D_P$ is the rate of diffusion, $c$ is the migration rate due to information aggregation, and $g(y)$ is the interaction function (perception kernel function) taken over a region $R$ (in one dimension, this is taken to be the radius of interaction). The migration term is nonlocal allowing each position in space to sample the state of opinions all around. The nonlocal interaction property plays the role of increased access to information. Mathematically, it is represented by the cross correlation between the opinion distribution and the interaction function. It can be interpreted as the total force on an individual subject to all the nearby populations. Sampling the opinions far away from your opinion is as easy as a web search revealing articles, videos, and more types of media available. The population density $P(x)$ at opinion $x$ is influenced by nearby populations in the opinion space, each of which is weighted by the perception kernel function $g(y)$. Alternatively, it can be thought of as the range of influence the population with opinion $x$ has on its neighbors.

It is interesting to note that Eq.\ \eqref{eq:model} can be reduced to the following form of a conservation law
\begin{align}
    \pdv{P}{t} + \nabla \cdot J = 0,
    \label{eq:conserve}
\end{align}
where $J$ is the population flux accounting for both diffusion and migration toward nonlocal choices,
\begin{align}
    J = -D_P\nabla{P} + cP\int_{R}P(x+y,t)g(y)dy.
\end{align}
The significance here is that the model assumes a constant total population. This modeling decision is common when studying a predefined social network from data, however it may not accurately account for the fact that populations fluctuate.

\subsection{Pattern Forming Instability with Biased Interaction Kernel Function} \label{sec:linear-stab}

We consider a form of the perception kernel function $g(y)$ that can be either unbiased, moderately biased, or heavily biased. An odd-symmetric perception kernel function represents a fully unbiased population that considers information equally from the right and from the left. Human actors are not always unbiased, so we introduce a perception kernel function with a bias parameter $b\in[0,2]$. The range outlines representations from completely unbiased ($b = 0$) to heavily biased ($b=2$). The functional form of the kernel function is given by 
\begin{align}
    g(y) = e^{-\left(\frac{y-\mu}{\sigma}\right)^2} + (b - 1) e^{-\left(\frac{y+\mu}{\sigma}\right)^2},
    \label{eq:newKernel}
\end{align}
and shown in Figure \ref{fig:kernels}. The parameters $\mu$ and $\sigma$ are included for generalization of enhanced information gathering similar to Sayama \cite{sayama2020enhanced}. Without losing generality of our result, we choose to keep these fixed at $\mu = 1$ and $\sigma = 1$. Note that this choice in parameters does induce a dimensionality in terms of $\sigma$. For instance, $x\rightarrow x/\sigma$ and $t\rightarrow t/\sigma$. Incidentally, this does not change the qualitative results we observe from numerical simulation thus we forwent a dimensional reduction analysis.

The interesting part of Eq.\ \eqref{eq:newKernel} is that bias introduces dynamics such as self-propelling motion and aversion (repulsion) that is not often integrated in opinion dynamics models. By `self-propulsion' we mean that the interaction kernel function $g(y)$ is non-zero at the origin. This results in the advection term in Eq.\ \eqref{eq:model} (second term) will cause motion simply due to a population interacting with itself. This may be that some populations may want to change their opinion regardless of the state of opinions nearby. 

The spatial dimension in Eq.\ \eqref{eq:model} represents an opinion on a continuous scale. The interpretation of position $x$ could be political, for instance, where a person lands relative to the right-wing to left-wing spectrum. Alternatively, we can think of the spatial position representing the magnitude of eccentricity assigned to an opinion, $x=0$ being standard or uninteresting opinions and $x=L$ being the most extreme or eccentric opinions. Then, a drift to the right would capture a population's preference toward more eccentric or extreme opinions.

To study the onset of pattern formation in Eq.\ \eqref{eq:model}, consider an infinitesimal perturbation along the spatially extended direction (here to be the $x$ direction) to a homogeneous state $P_h$. Substituting $P(x,t) \rightarrow P_h + \Delta{P(t)} e^{ikx}$ assumes periodic structures and also alleviates the model of spatial derivatives. This allows us to derive a linear model in terms of the amplitude of perturbation $\Delta{P}$. We choose the shape of the perturbation to be a complex amplitude to achieve the most general result in the linear model.

\subsection{Local Approximation}
Dynamically speaking, information aggregation serves as an attractive force between regions in the opinion space. This effectively models the cooperation aspect in social systems. The initiation of group formation depends on the shape of the biased interaction kernel function $g(y)$. This is not inherently clear from the nonlocal model Eq.\ \eqref{eq:model}. We seek to decompose the nonlocal term to recover local dynamical properties.

The nonlocal information aggregation term refers to the following
\begin{align}
    \int_R P(x+y,t) g(y) dy.
\end{align}
Assume that the region of interaction $R$ is small compared to the length of the opinion space. Therefore, $y$ is also very small, so we can Taylor expand the distribution function (dropping the $t$ for clarity)
\begin{align}
    P(x+y) &\approx P(x) + \partial_xP(x)y + \frac{1}{2!}\partial_x^2P(x)y^2 + \ldots
\end{align} 

Making this substitution in the model allows us to factor out the dependence on the distribution $P(x,t)$ in the integral. Inside the integral will then contain a series of moments of the interaction kernel function, terms of the form $y^n g(y)$. At the same time the population distribution is made local. The model then contains an infinite series of higher order derivatives. As an approximation, the series is truncated within sufficient enough terms. The number of terms we will use in this approximation is two 
\begin{align}
    P(x+y) \approx P(x) + \partial_xP(x)y,
\end{align}
for a baseline look at the local representation of population aggregation.

\subsection{Numerical Simulation and Measurements of Dynamics} \label{sec:measures}
To test the results of linear stability analysis, we conduct a series of numerical simulations. We used a standard method to discretize the continuous space into $N$ equally sized cells, and for the temporal evolution we used a forward-Euler algorithm. Simulating PDEs can lead to numerous numerical complications. To combat this, we conducted a series of trial and error test scenarios to achieve spatial and temporal resolutions, $\Delta{x}$ and $\Delta{t}$, that produced qualitatively robust behavior. The simulations are conducted in the Julia 1.8.2 programming language using only the base libraries. Source code is available online \footnote{Simulation code at : https://github.com/ckoertje/PDE-Opinion-Dynamics-w-Bias/}.

PDEs of pattern forming nature tend to have a dependency on the ratio between the length of domain and spatial resolution. Our simulations maintain the same length $L$ of the opinion space and same resolution $\Delta{x}$. This is so we are studying the effect of perception kernel function shapes. We also keep the ratio of the diffusion rate and migration rate $D_P/c$ constant, and we only vary the bias parameter $b$ in the kernel function. 

Here we use the temporal correlation function to capture the temporal change in the popularity distribution given the different amounts of bias in information gathering. The temporal correlation function is defined as 
\begin{align}
    C(x,\tau) = \left<P(x,t)P(x,t+\tau)\right>_t - \left<P(x,t)\right>_t^2,
\end{align}
where the angled brackets indicate an average over $t$. We begin the measurement of temporal correlation sufficiently long after initial groups are formed which we set to be $\tau = 0$. To avoid picking any one spot, we calculate the correlation at each position in the opinion space and then take the average $\left<C(\tau)\right>_x$ this time over $x$. This results in one value for the correlation after time $\tau$. We also normalize the correlation by dividing through by the initial value $\left<C(\tau = 0)\right>_x$.

The temporal correlation function allows us to measure changes in the opinion distribution through time and measure characteristic time scales for the motion of collective groups. We expect the steady state solutions, those of which patterns don't change, to have a one-to-one correlation over the course of the simulation. This indicates that the population distribution remains in the same ordered configuration during the course of the simulation. We run multiple simulations for each preference setting to improve accuracy of the result. It is also possible that spatial defects like dislocation (merging) of two groups may occur long after initial patterns are formed. This will have a minor impact on the correlation.

In addition to the temporal correlation, we define another measurement through the use of distance functions. We wish to characterize how different a distribution is to a future distribution at regular time intervals $t\rightarrow t+\Delta$. If the population reaches a steady state, then the distance between subsequent measured distributions will be zero (or very close to zero if merging occurs). If the population returns to the stable homogeneous state, the total amount of changes will also go to zero, however, in between these extremes there will be a peak at which the distribution is most different in that particular time interval. The first distance we use is the Manhattan distance (or L$^1$-norm) defined as, 
\begin{align}
    d_M = \int_0^L \left|P(x,t) - P(x,t+\Delta)\right| dx.
\end{align}
The other distance is the Euclidean distance (or L$^2$-norm), 
\begin{align}
    d_E = \sqrt{\int_0^L \left[P(x,t) - P(x,t+\Delta)\right]^2 dx}.
\end{align}
After collecting data for many bias parameter $b$ settings, we average the distances over time to achieve a distance measure as a function of $b$.

\subsection{Impact of Boundary Conditions on Unbiased Group Formation}
By abstracting the opinion values into a continuous space, we must also consider the structure of the space and its bounds. For the simulations of the bias interaction model, we are making use of periodic boundaries for ease of simulation. However, opinions may not exist on a circle, but rather they may go on infinitely in either direction or they may be bounded by a hard wall. 

The scenario with infinite boundary conditions is interesting in terms of extremism purposes because groups will drift forever, however it is easy to see the dynamics will not be tremendously different from the periodic scenario. 

Here we explore the situation in which the opinion space has a defined size and the boundary values are of Dirichlet type. We will give the distribution a value of zero on the boundaries to indicate that individuals cannot have opinions outside a predefined region
\begin{align}
    P(0,t) = P(L,t) = 0.
\end{align}
This change in boundary conditions will have an impact on pattern formation. In particular, large gradients are expected to form on the boundaries at the start of the simulation. Numerical simulations are conducted to study the effect hard boundary conditions have qualitatively on the formation of opinion groups.

\section{Results}
\begin{figure}
    \centering
    \includegraphics[width=0.99\columnwidth]{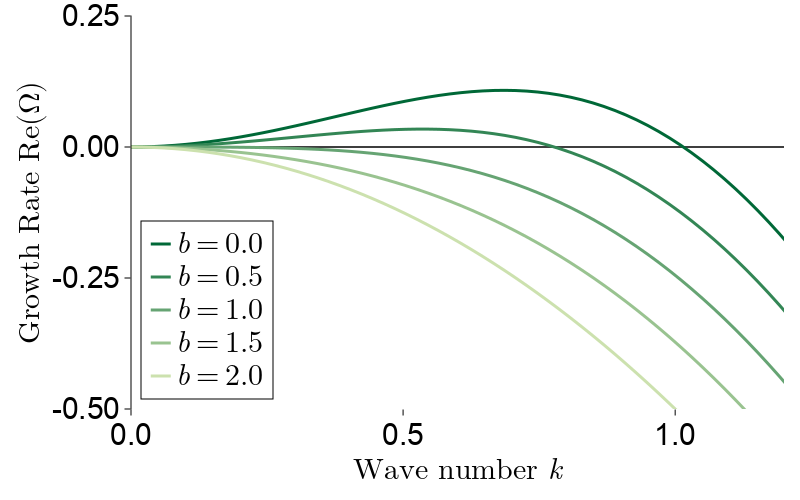}
    \caption{A plot of the real part of the dispersion relation from Eq.\ \eqref{eq:result-ls} for a variety of values for the bias parameter $b$ keeping the ratio $D_P/c = 1/2$. Unbiased populations ($b = 0.0$) have the largest possible wave number leading to the smallest possible characteristic wavelength. As the population becomes more and more bias, the largest wave number decreases and the characteristic wavelength will increase leading to greater polarization. In this case, a bias value of $b=1.0$ is the transition point.}
    \label{fig:rate}
\end{figure}
\subsection{Pattern Formation Instability with Biased Interaction Kernel Function}
Keeping only linear terms of $\Delta{P}$ and setting $P_h=1$, the resulting linear model is given by
\begin{align}
    \dv{\Delta P}{t} = \left(-D_P k^2 - ick\int g(y)dy - ick\int e^{iky}g(y)dy\right)\Delta P.
\end{align}
Details of this derivation are in Appendix \ref{app:linear-stability}. The growth rate $\Omega$ of the perturbation is a function of the spatial wave number and corresponds to the term inside the parenthesis
\begin{align}
    \Omega(k) = -D_P k^2 - ick\left(\int g(y)dy + \hat{g}(k)\right),
    \label{eq:rate-eq}
\end{align}
where $\hat{g}(k)$ is the Fourier transform of the perception kernel. 

Given Eq.\ \eqref{eq:rate-eq}, the dispersion relation depends on the shape of the perception kernel $g(y)$ and its Fourier transform. Assuming the perception kernel $g(y)$ is a real valued function, the Fourier transform can be decomposed into a real and imaginary part,
\begin{align}
    \hat{g}(k) = \int g(y) \cos{(ky)}dy + i\int g(y) \sin{(ky)}dy.
    \label{eq:fourier}
\end{align}
Doing so identifies that the amplitude growth rate can have a purely real and purely imaginary component not previously reported \cite{sayama2020enhanced}. This is due to our choice in perturbation having a complex valued wave form. If the perception kernel $g(y)$ is odd symmetric for an unbiased population, then the cosine integral in Eq.\ \eqref{eq:fourier} goes to zero, and groups can form when the real part of $\Omega(k)$ (Eq.\ \eqref{eq:rate-eq}) is greater than zero. Conversely, if the perception kernel $g(y)$ is even symmetric for a heavily biased population, then the sine integral in Eq.\ \eqref{eq:fourier} goes to zero and the only real part in Eq.\ \eqref{eq:rate-eq} left is always negative. This means that a heavily biased population will never form groups.

The results of the linear stability analysis for our interaction kernel function yields the following dispersion relation 
\begin{align}
\begin{split}
    \Omega{(k)} &= - D_P k^2 + ck \frac{(2-b)}{2} e^{-k^2/2}\sin{(k)} \\ 
     &\qquad - ick b\left(\frac{1}{2}e^{-k^2/2}\cos{(k)} + \sqrt{\pi}\right).
\end{split}
\label{eq:result-ls}
\end{align}
Figure \ref{fig:rate} is a plot of the real part of $\Omega$ as a function of the wave number $k$. The shape of this stability curve is consistent with other conservative systems \cite{cross1993pattern}. The transition from stable to unstable starts with small wave numbers first. This emits large wavelength patterns,
\begin{align}
    \lambda = \frac{2\pi}{k}
\end{align}

Lightly biased populations have more available wave numbers with positive growth rates meaning they can form groups more readily. A population that is too biased will no longer be able to form groups as the homogeneous solution becomes stable. In this example, the unbiased ($b=0.0$) actors will self-organize according to the shortest possible wave length between choices, $\lambda = 2\pi/k_{max}$. At this point the distance between groups will be the smallest. As we increase the amount of bias $b$, then the largest wave number becomes smaller and the characteristic wave length will grow. 

While the dispersion relation reveals whether pattern formation is available to the social system, we seek an expression for what bias setting no longer facilitates group formation; we will call this the ``critical bias'' threshold/setting. We can do so by the neutral stability analysis. Neutral stability is defined by the boundary between the stable and unstable regions of the model where Re$(\Omega) = 0$. This leads to an expression for the bias as a function of the wave number $b(k)$. The critical bias then corresponds to the value at which the last wave number transitions from unstable to stable. For a conservative system like the model here, this corresponds to $b(k\rightarrow0)$.

\begin{figure}
    \centering
    \includegraphics[width=0.99\columnwidth]{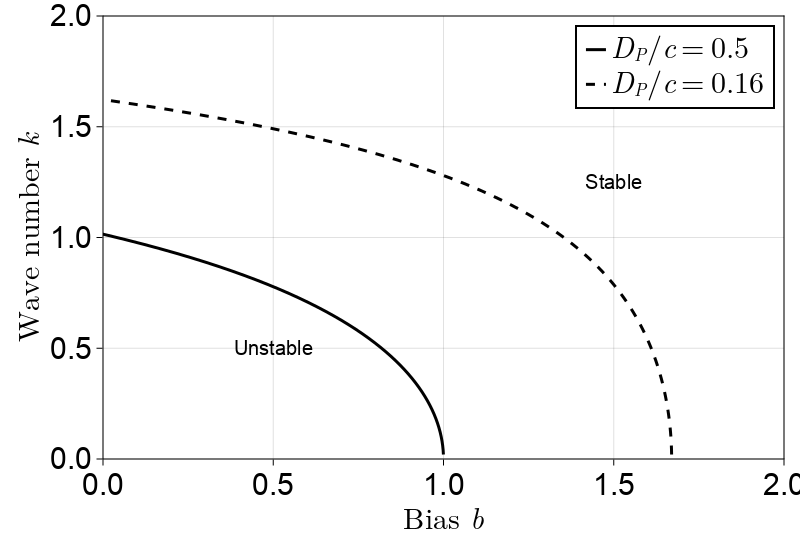}
    \caption{Plot of two boundaries showing the regions of stability and instability for wave numbers over the range of bias values $b$. The boundary is plotted from Eq.\ \eqref{eq:bound}. As the ratio $D_P/c$ decreases, or as the impact of diffusion becomes less and less strong, the region of possible pattern formation (corresponding to the unstable region) grows.}
    \label{fig:phase}
\end{figure}

There is a balance maintained by the diffusion and migration. This is determined by the ratio of the diffusion rate and the migration rate $D_P/c$. From Figure \ref{fig:phase}, populations in which mutual interactions are more impactful (ratio $D_P/c$ decreases) are more readily eager to form groups given for any level of bias. This is reflected in how the boundary shifts to the right. The equation of the boundary comes from the neutral stability analysis and takes the form
\begin{align}
    b(k) = 2 - \frac{2D_P}{c} k e^{k^2/2} \csc{k}.
    \label{eq:bound}
\end{align}
The details of this derivation are found in Appendix \ref{app:linear-stability}. This is defined over the region $b\in[0.0,2.0]$ representing the range of an unbiased population to a heavily biased one.

We can calculate the transition point as the critical bias setting that determines stability. The critical bias setting is achieved by taking the limit of Eq.\ \eqref{eq:bound} as $k\rightarrow0$. The result takes the form
\begin{align}
    b_c = 2 - \frac{2D_P}{c}.
\end{align}
For example, when $D_P/c = 0.5$, the critical transition point is $b_c = 1.0$ for a moderately biased population. This matches the numerical result in Figure \ref{fig:rate}. Details of this limit derivation are found in Appendix \ref{app:transition}. Here, the dynamics now boil down to two possible regions. That is pattern formation where $b/b_c < 1$ and no pattern formation where $b/b_c > 1$.

\begin{figure*}
    \centering
    \includegraphics[width=0.99\textwidth]{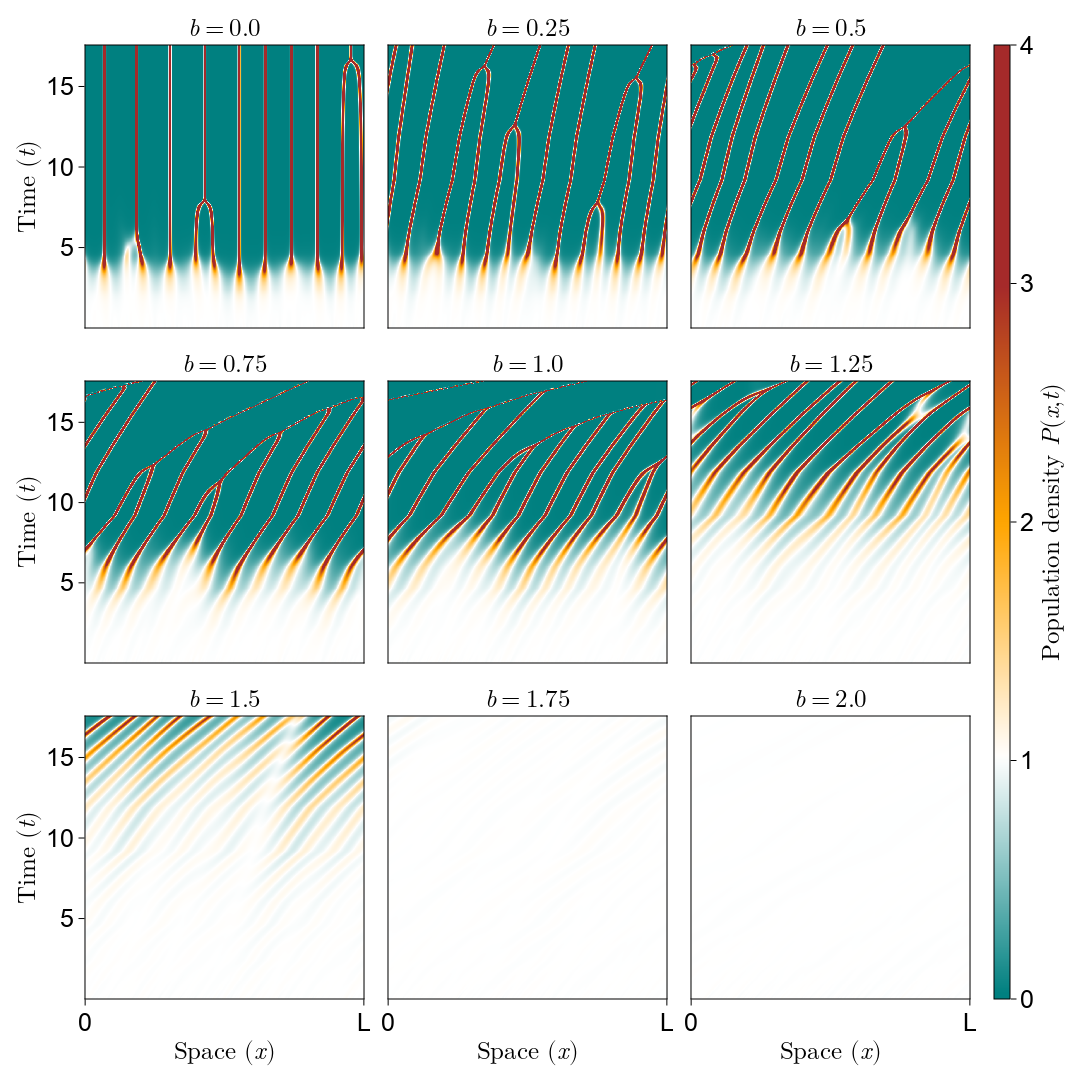}
    \caption{An array of simulation runs by varying the bias parameter $b$ and keeping the ratio $D_P/c = 0.16$. This choice of parameters is to show more slowly how the group motion tilts before the homogeneous solution becomes stable at $b/b_c > 1$. Here, the length of the opinion domain $L = 62$, spatial resolution $\Delta{x} = 0.062$, and temporal resolution $\Delta{t} = 0.001$. When $b=0.0$, the complex component of the dispersion relation is 0 and thus the resulting patterns are stationary. On the other end when $b=2.0$, there are no positive real valued growth rates for any wave number resulting in a uniform distribution with no patterns. In simulations between these extremes, we observe a collective group drift towards one end of the domain. The transition point for this simulation occurs at $b=1.68$ in alignment with Figure \ref{fig:phase}.}
    \label{fig:many}
\end{figure*}

\subsection{Local Approximation}
The result of localizing the non-local term leads to a new PDE model with local dynamics 
\begin{align}
    \pdv{P}{t} &= D_P\partial_x^2{P} - c \partial_xP \sum_{n=0}^{\infty} \frac{a_n}{n!} \partial_x^{n} P - c P\sum_{n=0}^{\infty} \frac{a_n}{n!} \partial_x^{n+1} P
\end{align}
where $a_n$ refers to the n$^{th}$ moment of the bias interaction kernel function $g(y)$. The details of the derivation are found in Appendix \ref{app:local}. To gain some insight from this new model, consider only the terms up to $n=1$ 
\begin{align}
    \pdv{P}{t} &= D_P\partial_x^2{P} - 2ca_0P\partial_xP - ca_1 \left(\partial_xP^2 + P\partial_x^2 P\right).
\end{align}
Population aggregation behaves like the last three terms. Of particular interest, the last term is nonlinear diffusion with a negative coefficient. Anti-diffusion seems to be a mechanism for opinion dynamics. When one opinion is more popular than others, the population will seek to aggregate around it. Similar mechanisms are found in approximations of the Deffuant model \cite{alexanian2018anti}. To achieve pattern formation in this localized model, we would suggest including terms atleast up to a 4$^{th}$ order spatial derivative $n=3$.

\subsection{Numerical Simulation and Measurements of Dynamics}
Simulating the model outright with the new perception kernel reveals new dynamic patterns. For kernels that are no longer perfectly unbiased, the resulting groups experience a collective drift motion during and after group formation; see Figure \ref{fig:many}. Lightly biased groups drift to the right toward more eccentric opinions very slowly, however with greater bias, the speed of the drift increases. This is evident in the amount of 'tilt' or the angle in spacetime that increases with greater bias. Interesting to note, more popular groups (taller peaks) tend to drift faster due to the nonlinearity in the advection term in Eq.\ \eqref{eq:model}. As a result, if the simulation were left to continue for a much longer time, we would see all the peaks coalesce into one group due to the periodic boundary conditions. Given infinite boundary conditions, some groups may end up getting left behind to the left while larger groups drift faster away to the right. The critical transition point in the simulation in Figure \ref{fig:many} is a bias value $b=1.68$.

\begin{figure}
    \centering 
    \includegraphics[width=\columnwidth]{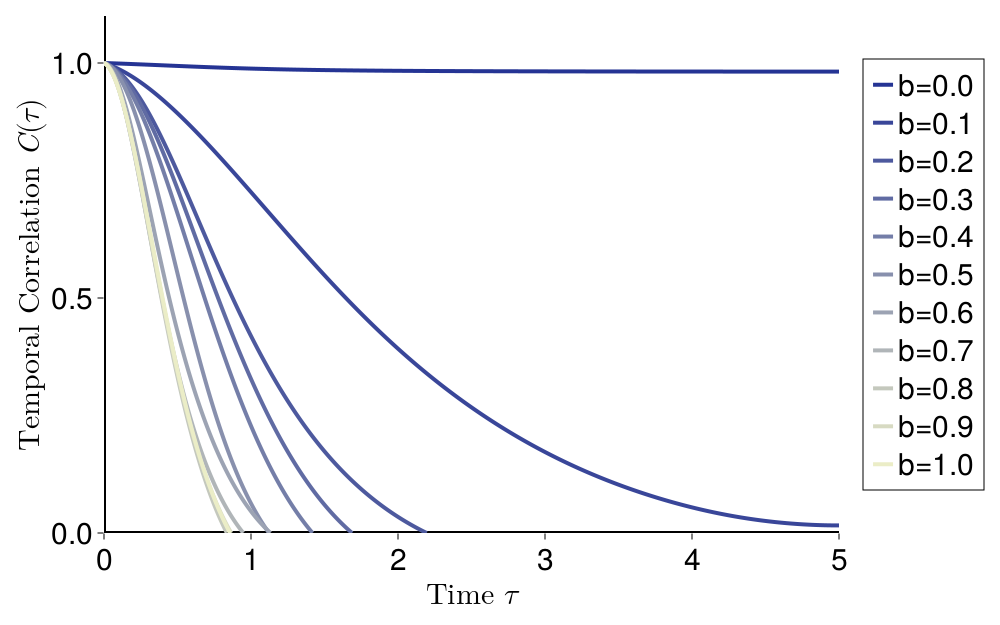}
    \caption{Measurements of the average temporal correlation function ranging from an unbiased population $b=0.0$ to a moderately biased population $b=1.0$. In this experiment ($D_P/c = 1/2$), $b=1.0$ is a transition point, so we do not include the simulations that return to the homogeneous state. The values plotted are those up until the system loses temporal correlation. The more biased a population is the quicker the drop off.}
    \label{fig:corr}
\end{figure}

Figure \ref{fig:corr} shows the characteristic time for biased populations to lose temporal correlation. A fully unbiased population remains in roughly the same configuration through time as seen when $b=0.0$, so the temporal correlation stays at 1. However, introducing a small amount of bias will cause the population to change its order slowly. More and more bias causes the population to change up opinions more and more quickly. This captures the average speed at which a population will drift to more eccentric opinions. These dynamics appear to be understudied in realm of opinion dynamics.

\begin{figure}
    \centering
    \includegraphics[width=\columnwidth]{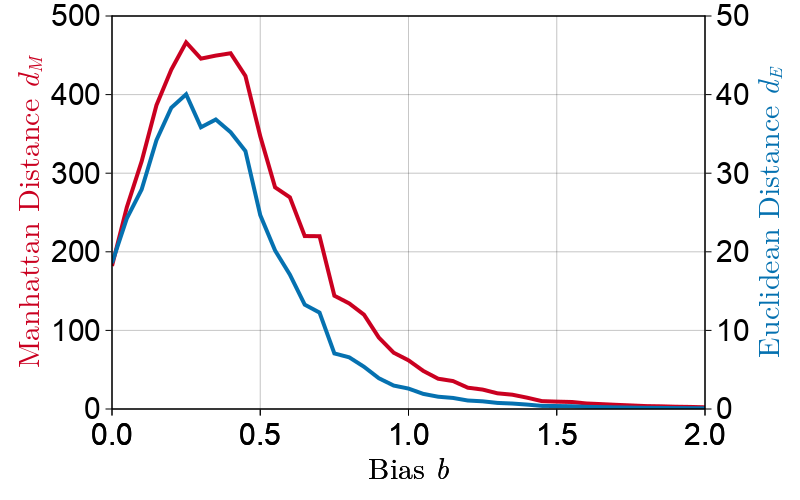}
    \caption{Measurements of the Manhattan distance (blue) and Euclidean distance (red) and Manhattan distance (blue) on average from time $t\rightarrow t+1$ for various levels of bias. The simulation parameters are the same for the numerical simulations in Figure \ref{fig:corr} to get a profile that was not too heavily skewed. We measured the distance between 10 different time steps and 5 separate simulations (totaling to 50 measurements for each point in the graph). Increasing the number of simulations will smooth out the curve.}
    \label{fig:distances}
\end{figure}

When considering consistent time periods $t\rightarrow t+1$, populations with a small amount of bias, $b \approx 0.25$ in particular, appear most different; see Figure \ref{fig:distances}. The bias setting for the population that changes the most is subject to vary depending on the time interval we choose. Say we survey a population once a year, then we can measure how different the opinions are through time and map it to how biased the population ought to be. By modifying the time between sampling, the peaks in Figure \ref{fig:distances} will shift right or left based on allowing more or less time respectively for the population to drift. An approximate analytical scheme for obtaining this bias setting may be subject of future investigation and mapping data could help connect this result to the real world. The time scale from $t \rightarrow t+1$ is what takes a population of that level of bias to drift from their initial group position to the position of a new opinion that was previously unpopular. Populations with greater bias become stable to homogeneous solutions so the distribution of opinion population through time would not change. The initial mixed opinion state stays mixed throughout time. The significance of this measurement is that if you sample a population with surveys on a regular basis, you may be able to deduce how biased a population is.

\subsection{Impact of Boundary Conditions on Unbiased Group Formation}
\begin{figure}
    \centering
    \includegraphics[width=\columnwidth]{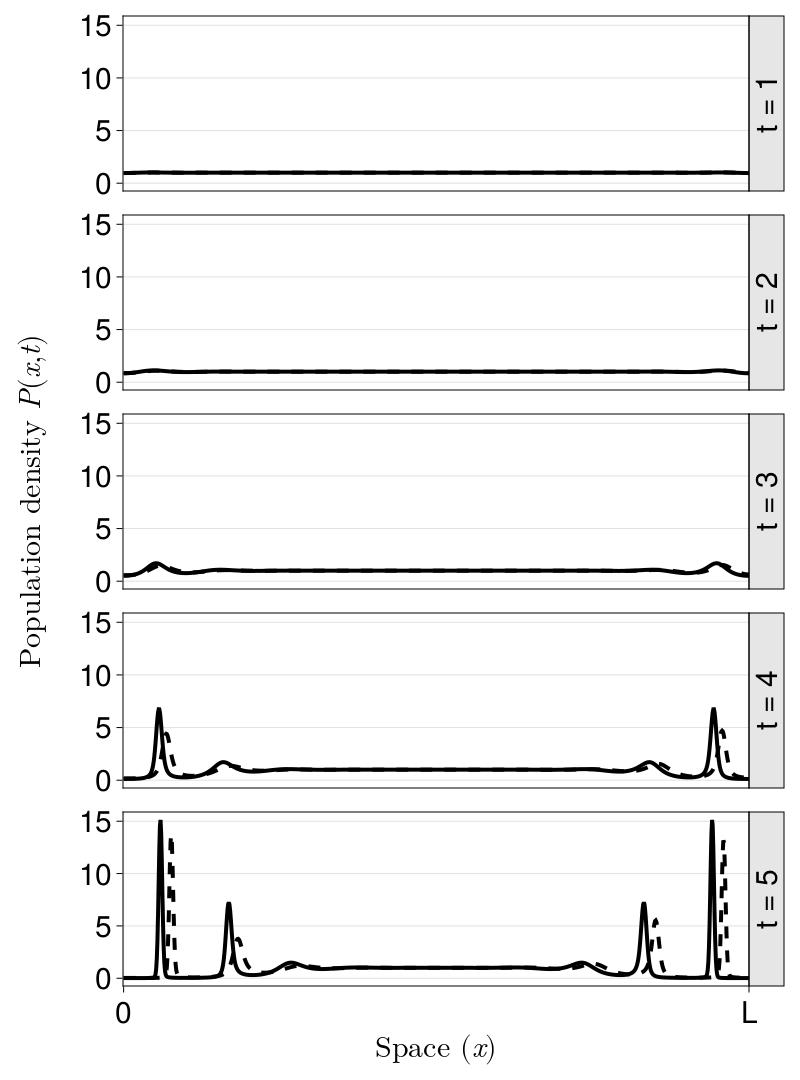}
    \caption{A series of population density distributions at different times simulated with Dirichlet boundary conditions $P(0,t) = P(L,t) = 0$. The solid black line is for $b=0$ and the dashed line is for $b = 0.4$. Initial pattern formation occurs towards the edges of the opinion space and propagate inwards.}
    \label{fig:hardbc-state}
\end{figure}

When we place fixed boundaries on the opinion space, those on the edges have fewer opinions to interact with on one side. What we observe is that groups quickly form toward the most extreme sides of the space (Figure \ref{fig:hardbc-state}). Giving a mixed population, say for a case study, a new topic to debate on, the most definitive opinions form first due to the large gradients generated by the boundaries. Essentially, this represents the definitive 'strongly agree' or 'strongly disagree' groups that form. After that, as those in the middle continue deliberating, we start to see moderate groups forming towards the center.

With bias in the population, groups no longer have the option to wrap around the domain. As a result, groups begin to pile up on the right end of the space. Eventually, all participants come to an agreement in extremity. We do not observe any kind of rebound dynamics because the population as a whole keeps its preference toward opinions on the right.

We did not test free boundaries conditions in this study, but we expect the same qualitative behavior as the periodic boundary simulations. The difference there is that the opinion groups can drift outside of our window on forever. Another difference is that instead of the biased groups all coalescing into one, large groups may be able to escape from slower moving groups. The question that then comes up is what it means for the opinion domain to be infinite. Extending to higher dimensions may be a better approach to this.

\section{Discussion}
Modeling social systems at different scales not only helps in understanding the systems themselves, but also in tackling the fundamental properties of self-organizing dynamics. We examined two modeling extensions and applied them to a PDE-based model of opinion dynamics. Those were biased information gathering and the effect of boundary condition on pattern formation. The partial differential equations approach seeks to study the problem of how people form opinions at a macroscopic description differing from those studied in complex social networks. This is especially beneficial when considering large populations of which constructing an explicit social network is impossible and computationally demanding. From the analysis presented here, bias (modeled as an asymmetric dynamical interaction) may to be a mechanism behind the increased interest in the formation of extreme groups. 

Even small amounts of bias induces a collective group drift in the resulting groups such that the distribution slowly moves towards the eccentric end of the spatial domain. The linear and neutral stability analyses demonstrate the potential for pattern formation given parameter settings along with the critical value for the bias parameter where group formation becomes no longer possible. Numerical simulations show that as groups drift, the more popular opinions tend to drift faster and merge with smaller groups. On average, due to this motion, the organization of the opinion distribution will change at different rates. The temporal correlation function showed how the speed at which distributions change depends on the amount of bias in the population. More biased populations will deviate more rapidly. Through use of a periodic distance measurement, we may be able to determine how bias a population is.

By treating opinion dynamics as a spatial system, the boundaries to the space become important for the dynamics to unfold. We found that fixed boundaries force opinion formation more readily on the boundaries before moderate opinions are formed. This results in a greatly polarized population at the beginning. Due to the characteristic distance desired by the model, it is also observed that the groups on the edges are slowly forced closer and closer to the extreme boundaries. Boundary conditions may be enough to push groups out to the extreme ideologies.

The collective group drift is particularly interesting in that we may be seeing large masses of people change their opinion in sync towards an eccentric direction. This could be related to discussions of increased polarization by having two peaks that drift in opposite directions. We did not observe these dynamics in the model presented here because the bias perception kernel function is only bias toward one direction. In reality, the shape of the perception kernel function may depend on the position in the domain.  That way the distance in opinion space between the groups would increase through time. Boundary conditions play a role in limiting how far they can go. We utilized periodic boundary conditions to get a sense of long term behavior, however we note how the structure of opinion spectrums may not be periodic in nature. For instance, they may be infinite, in which case, the drift would go on forever toward greater and greater extremism. High dimensions may help to understand the meaning of this scenario. Also, due to the nonlinearity in migration speed, it may be possible that some smaller groups are left behind. It may also be possible to have hard boundaries from which drifting groups rebound off of and drift back in the opposite direction much like active particle dynamics. Again, this is not observed in our model because bias is geared towards one direction only.

We also note how populations in the real world do not remain constant as the current model assumes. For instance, Eq.\ \eqref{eq:model} follows a conservation law implying that the population in the opinion domain is constant. The model here may represent an already predefined social network whose interactions can drive this particular group to develop extreme ideas. Including a simple population growth reaction term may be sufficient for unique dynamics like spatio-temporal chaos. We previously studied population kinetics through a logistic model. This reaction term alone causes the peaks to be squashed by overgrowth and the troughs to grow seemingly out of nothing \cite{koertje2023modeling}. It is anticipated that additional reaction terms will lead to additional spatiotemporal dynamics of interest.

More research should be focused on heterogeneous populations to understand the effect of mixing different types of people. This could be by letting the interaction kernel function depend on the location in space $x$. That way each position is subject to a different interaction kernel function representing varying degrees/directions of bias in one simulation. The opinion dynamics scene needs more modeling efforts into non-static pattern formation as shown here.

Due to the stochastic nature of human behavior, it may be useful to relate the model studied here to statistically motivated models such as the kinetic model from Toscani \cite{toscani2006kinetic}, where we note the formalism using Focker-Planck models. The local limit of the model studied here can be further studied to understand pattern formation in single species systems.

Another potential next study is to introduce some form of social impact gradient in the background to drive the system. The idea here is similar to a temperature gradient in a system like Rayleigh-B\'enard convection. Another way to drive the system might be with additional perturbations at regular time periods. This would represent public events that shock the population. 

One of the current issues in social dynamical modeling is its relationship with empirical data. Connecting this research with experimental data is a tough task. A first step is to find a mapping between our results to other models and real-world data. Within data science, there are techniques of semantic analysis based on social media reactions or comments that may have an insight into the shape and structure of an individual's perception of information \cite{pandey2023generation}. Without knowing the shape of the perception kernel, we can make assumptions and study the possible resulting behavior. The actual shape of the kernel may be determined experimentally through surveys or interpolated from data on social media.

A lot of experimental probes into opinion dynamics are also focused on social networks in how people interact. It may be beneficial to measure macroscopic trends to see how the distribution of opinions may change. Related to this study, it would be to measure the temporal correlation function or the distribution distances from experimental data. One approach could be to quantify how eccentric political opinions have changed over a few decades. In the U.S.\, for instance, we have seen political majority deviate every few years based on presidential elections.

Modeling complex human interactions through continuous opinion dynamics is a very flexible and robust way to study how people come to agreements and disagreements. We show here how adjustments to a simple model can lead to new dynamics not typically discussed in opinion dynamics. This leads to new potential directions to explore for trying to understand social interactions. Continuing to improve this understanding will lead to better social policies that limit extreme or dangerous group formation.

\appendix
\section{Linear Stability Details with Biased Interaction Kernel}\label{app:linear-stability}
Starting with the model equation in one-dimension
\begin{align}
    \pdv{P}{t} = D_P\pdv[2]{P}{x} - c\pdv{}{x}\left(P \int P(x+y)g(y)dy\right),
\end{align}
we make a small amplitude, time-dependent perturbation to a homogeneous solution. This leads to the substitution $P \rightarrow P_h + \Delta{P}(t) e^{ikx}$. The expansion comes out to 
\begin{widetext}
\begin{align}
\begin{split}
    e^{ikx}\dv{\Delta{P}}{t} &= -D_Pk^2 \Delta{P} e^{ikx} - ick \Delta{P} e^{ikx}\int(P_h  + \Delta{P} e^{ik(x+y)})g(y)dy - ick(P_h + \Delta{P} e^{ikx})\int \Delta{P} e^{ik(x+y)} g(y)dy,
\end{split} \\
\begin{split}
    e^{ikx}\dv{\Delta{P}}{t} &= -D_Pk^2 \Delta{P} - ick P_h \Delta{P} e^{ikx} \int g(y) dy - ick P_h \Delta{P} e^{ikx} \int e^{iky}g(y) dy,
\end{split} \\
    \dv{\Delta{P}}{t} &= \left(-D_Pk^2 - ickP_h \left[\int g(y) dy + \int e^{iky}g(y) dy\right] \right)\Delta{P}.
\end{align}
\end{widetext}
In the second step, we eliminated any term containing order $\Delta{P}^2$ as $\Delta{P}$ is already small. We eliminated the exponential from each term to simplify further. The homogeneous steady state $P_h$ is arbitrary and can be set equal to 1 to clean up the result. The end result is a linear model in terms of the amplitude perturbation with growth/decay rate as a function of $k$,
\begin{align}
    \Omega(k) = -D_P k^2 - ick\left(\int g(y)dy + \int e^{iky}g(y)dy\right).
    \label{eq:disp}
\end{align}

 An analytical form of Eq.\ \eqref{eq:disp} can be obtained through the use of our particular perception kernel
 \begin{align}
     g(y) = e^{-(y-1)^2} + (b-1)e^{-(y+1)^2}.
 \end{align}
 We chose $\mu = \sigma = 1$ to simplify the analysis. Note this choice induces a dimensionality of $x/\sigma$, $t/\sigma$, etc. Dropping this dimensional reduction rigor does not change our results qualitatively. The analytical form can be achieved by performing the integrals in Eq.\ \eqref{eq:disp}, assuming the bounds are infinite for ease of calculation
 \begin{align}
     \int_{-\infty}^{\infty} g(y) dy &= b\sqrt{\pi}.
 \end{align}
 The second integral is simply the Fourier transform of the perception kernel $g(y)$. The result after some manipulation is 
 \begin{align}
     \hat{g}(k) = \frac{b}{2} e^{-k^2/2}\cos{(k)} + \frac{(2-b)}{2}ie^{-k^2/2}\sin{(k)}.
 \end{align}

 All together, the result of the linear stability analysis yields the dispersion relation
 \begin{align}
 \begin{split}
     \Omega{(k)} &= - D_P k^2 + ck \frac{(2-b)}{2} e^{-k^2/2}\sin{(k)} \\ 
     &\qquad - ick b\left(\frac{1}{2}e^{-k^2/2}\cos{(k)} + \sqrt{\pi}\right).
 \end{split}
    \label{eq:app-rate}
 \end{align}
 Plots of the real part for different values of $b$ are shown in Figure \ref{fig:rate}.

We can then use the result from the dispersion relation to find curves of neutral stability. Since we only care about the real part for stability purposes, set the Re$\Omega(k)=0$ and solve for $b$. The result is,
\begin{align}
    b(k) = 2 - \frac{2D_P}{c} k e^{k^2 /2} \csc{k}.
    \label{eq:app-bound}
\end{align}
A plot of the boundary along with stable and unstable regions for given values of $b$ are shown in Figure \ref{fig:phase}.

\section{Derivation of Transition Point}\label{app:transition}
A point of interest is at what bias value the dynamics of a homogeneous solution become stable. From the plots of the neutral stability curves in Figure \ref{fig:phase}, a population that is too heavily bias will no longer form groups. This occurs at the intersection of the $b$-axis or when $k=0$.

The right-hand side of Eq.\ \eqref{eq:app-bound} is undefined at $k=0$, so we take the limit as $k\rightarrow 0$,
\begin{align}
    b_c &= \lim_{k\rightarrow 0}\left(2 - \frac{2D_P}{c} \frac{k e^{k^2 /2}}{\sin{k}}\right) \\
    &= 2 - \frac{2D_P}{c} \lim_{k\rightarrow 0} \frac{k e^{k^2 /2}}{\sin{k}}.
\end{align}
The limit can be computed by L'\^{H}opital's rule and found to go to 1. Therefore the critical/transition point occurs at 
\begin{align}
    b_c = 2 - \frac{2D_P}{c}.
\end{align}

\section{Derivation of Local Approximation}\label{app:local}
The non-local information aggregation term refers to the following
\begin{align}
    \int_R P(x+y,t) g(y) dy.
\end{align}
Assume that the region of interaction $R$ is small compared to the length of the opinion space. Therefore, $x+y$ is very close to $x$, so we can Taylor expand the distribution function (dropping the $t$ for clarity)
\begin{align}
    P(x+y) &\approx P(x) + \partial_xP(x)y + \frac{1}{2!}\partial_x^2P(x)y^2 + \ldots
\end{align} 
With this substitution, the non-local term is approximated at the following
\begin{widetext}
\begin{align}
    \int_R P(x+y,t) g(y)dy &\approx
     P(x)\int g(y)dy + \partial_xP(x)\int yg(y)dy  + \frac{1}{2!}\partial_x^2P(x)\int y^2g(y)dy + \ldots  \\
     &= a_0P(x) + a_1\partial_xP(x)  + \frac{a_2}{2!}\partial_x^2P(x)+ \ldots  \\
    &= \sum_{n=0}^{\infty} \frac{a_n}{n!} \partial_x^n P
\end{align}
\end{widetext}
where $a_n$ refers to the n$^{th}$ moment of the bias interaction kernel. Substitute this back in to the model Eq.\ \eqref{eq:model}, 
\begin{align}
    \pdv{P}{t} &= D_P\partial_x^2{P} - c\partial_x \left(P \sum_{n=0}^{\infty} \frac{a_n}{n!} \partial_x^n P\right), \\
    &= D_P\partial_x^2{P} - c \partial_xP \sum_{n=0}^{\infty} \frac{a_n}{n!} \partial_x^{n} P - c P\sum_{n=0}^{\infty} \frac{a_n}{n!} \partial_x^{n+1} P.
\end{align}

\begin{acknowledgments}
We would like to thank Austin Marcus for the insightful discussions and suggestions throughout the completion of this work. Hiroki Sayama thanks financial support from JSPS KAKENHI Grant Number 19K21571.
\end{acknowledgments}

\bibliography{references}

\end{document}